\documentclass[10pt,letterpaper,twocolumn]{article} 

\usepackage{ol2}
\usepackage[draft]{hyperref}
\usepackage{amsmath,amssymb}
\usepackage[english]{babel}
\usepackage[latin1]{inputenc}
\usepackage{doi}
\usepackage{graphicx}

\addto\captionsenglish{}

\begin{document}

\twocolumn[ 

\title{Quantum-enhanced micro-mechanical displacement sensitivity}

\author{Ulrich B. Hoff,$^{1,*}$ Glen I. Harris,$^2$ Lars S. Madsen,$^1$ Hugo Kerdoncuff,$^1$ Mikael Lassen,$^1$\\ Bo M. Nielsen,$^1$ Warwick P. Bowen,$^2$ and Ulrik L. Andersen$^1$ }

\address{
$^1$Department of Physics, Technical University of Denmark, Fysikvej bld. 309, 2800 Kgs. Lyngby, Denmark\\
$^2$Centre of Excellence in Engineered Quantum Systems, University of Queensland, St Lucia, Queensland 4072, Australia \\
$^*$Corresponding author: ulrich.hoff@fysik.dtu.dk
}

\begin{abstract}
We report on a hitherto unexplored application of squeezed light: for quantum-enhancement of mechanical transduction sensitivity in microcavity optomechanics. Using a toroidal silica microcavity, we experimentally demonstrate measurement of the transduced phase modulation signal with a sensitivity $-0.72(\pm 0.01)$\,dB below the shot noise level. This is achieved for resonant probing in the highly under-coupled regime, by preparing the probe in a weak coherent state with phase squeezed vacuum states at sideband frequencies.
\end{abstract}

]

\noindent
Originally spurred by the field of gravitational wave detection, considerable attention has been attracted to achieving high-sensitivity interferometric detection of mechanical displacements. Within recent years, such displacement sensing techniques have also found application for cantilever-based single spin detection~\cite{Rugar2004}, cavity optomechanical magnetometry~\cite{Forstner2012}, and the quest to reveal quantum effects in mesoscopic mechanical systems~\cite{Teufel2011,Safavi-Naeini2012}.
Fundamentally, the total measurement uncertainty of any optical displacement readout has contributions from two distinct noise sources~\cite{Caves1980}:
\emph{imprecision noise} in the form of photon shot noise of the detected probe light and \emph{quantum back-action noise} due to a stochastic radiation pressure force imparted on the mechanical resonator by vacuum fluctuations entering the setup. Using coherent states of light, imprecision noise sets a limiting sensitivity $\propto 1/\sqrt{N}$, with $N$ being the deployed number of photons, suggesting that arbitrarily high sensitivity can be achieved simply by increasing the optical power. However, as $N$ increases quantum back-action noise $\propto \sqrt{N}$ contributes more and more, and an optimal sensitivity, the so-called Standard Quantum Limit (SQL), is reached when the two noise sources contribute equally to the total measurement imprecision at the mechanical resonance frequency. Being a context-dependent limit, the SQL is not absolute and can in principle be surpassed, unlike the Heisenberg limit enforced by the Uncertainty Principle. In microtoroidal resonators, however, the high optical quality factor will cause the onset of \emph{dynamical back-action} well before reaching the SQL. This effect was first observed in such systems by Rohksari et al.~\cite{Rokhsari2005} and it has recently been shown that the resulting parametric instability can be suppressed using feedback techniques~\cite{Harris2012}.

It has been long known that the injection of squeezed light can be used to improve the phase sensitivity of an interferometric measurement~\cite{Caves1981} and this has been experimentally verified for both Mach-Zehnder~\cite{Xiao1987}, Sagnac~\cite{Eberle2010}, and large-scale gravitational-wave~\cite{LIGO2011} interferometer topologies.
In this Letter, we report on experimental demonstration of squeezed light-enhanced mechanical transduction sensitivity using a tapered fibre coupled microtoroidal resonator at room temperature. With this implementation of quantum-enhanced interferometry, we have extended the applicability of the technique to the regime of micromechanical oscillators. The potential benefits of using squeezed light are numerous, e.g.: (i) it improves the minimum achievable sensitivity at a given probe power, thus providing a way to effectively shift the onset of dynamical back-action that would otherwise mask the improved imprecision noise as the power is ramped up, (ii) it shifts the SQL to experimentally more feasible power levels, and (iii) quadrature anti-correlations can in principle be exploited to beat the SQL by using an optimized squeezing phase~\cite{Pace1993}.

\begin{figure}[h]
\centerline{\includegraphics{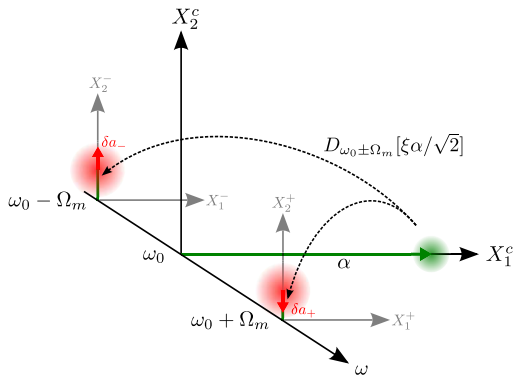}}
\caption{\label{fig_PhaseSpaceRep} Combined sideband and phase space representation of the optomechanical transduction mechanism. The sidebands are initially prepared in a composite two-mode squeezed state with the individual sidebands being in a thermal state. The optomechanical phase modulation transfers photons from the coherent carrier field (green) into the sidebands, displacing the initial quantum states $\delta a_{\pm}$ (red).}
\end{figure}

Micro toroidal resonators integrate a high-Q optical mode with a mechanical harmonic oscillator and couple the two via the radiation pressure mediated optomechanical interaction, formally described by the Hamiltonian $H_{om} = \hbar g_0 (b + b^{\dag})a^{\dag}a$. Here $a$, $a^{\dag}$ ($b$, $b^{\dag}$) are the optical (mechanical) ladder operators and $g_0$ is the vacuum optomechanical coupling rate. In the case of resonant optical excitation, the optomechanical interaction leads to pure phase modulation of the light field at the mechanical oscillation frequency $\Omega_m$. The modulation index is given by $\xi = g_0 \delta x/\Omega_m$, where $\delta x$ is the mechanical oscillation amplitude. In the sideband picture (Fig.~\ref{fig_PhaseSpaceRep}), this is formally equivalent to displacements $D_{\omega_0 \pm \Omega_m}[\xi \alpha/\sqrt{2}]$ of the $\omega_0 \pm \Omega_m$ sideband modes, where $\alpha$ is the coherent carrier amplitude. Thus the optomechanical interaction results in the excitation of weak coherent sideband states with amplitudes $\xi \alpha/\sqrt{2}$, carrying information about the mechanical oscillation frequency and displacement amplitude. This encoded information can be extracted by subsequent homodyne detection of the phase quadrature. The detection process is a joint measurement of the upper and lower sideband states at each detection frequency $\Omega$, and the resulting measurement imprecision noise, given by the spectral variance of the recorded phase quadrature, is~\cite{Glockl2005}
\begin{equation*}
V(X_2[\Omega]) \propto \beta^2 (V(X_2^+ + X_2^-) + V(X_1^+ -X_1^-))
\end{equation*}
where $\beta$ is the local oscillator field amplitude.
Thus, if the sideband states are initially vacua, the transduced signal will be measured with a noise equal to one shot noise unit. Alternatively, quantum correlated sidebands can be added symmetrically around the carrier, forming phase squeezed vacuum states. In this case the measurement imprecision noise is reduced below the shot noise level (SNL), yielding a quantum enhanced displacement sensitivity.

\begin{figure}[h!]
\centerline{\includegraphics{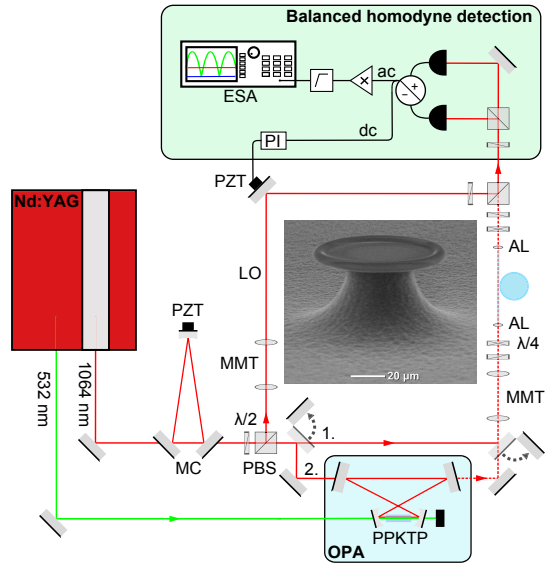}}
\caption{\label{fig_setup} Schematic diagram of the experimental setup. MC, mode cleaning cavity; PZT, piezoelectric transducer; PBS, polarizing beam splitter; MMT, mode matching telescope; LO, local oscillator field for balanced homodyne detection; AL, aspherical lens. Inset: SEM micrograph of the toroidal microcavity with major diameter $\simeq 60\,\mu\rm{m}$ and minor diameter $\simeq 6\,\mu\rm{m}$. At 1064\,nm the FSR is 1.04\,THz.}
\end{figure}

In the experiment, as shown schematically in Fig.~\ref{fig_setup}, light at $\lambda = 1064\,\rm{nm}$ from an Nd:YAG laser is first spectrally and spatially filtered by a mode cleaning cavity and subsequently split on a PBS to form probe and local oscillator fields. The probe field can follow two alternative beam paths: path 1 for coherent state probing and path 2 for generation of phase squeezed vacuum sideband states, in a PPKTP-based optical parametric amplifier (OPA)~\cite{Lassen2010}. Following a mode matching telescope, the probe field is coupled into a bare optical fibre (Corning SMF-28) using a single antireflection (AR) coated aspheric singlet lens with $f=8.07\,\rm{mm}$. For evanescent coupling to the micro toroidal resonator, the bare fibre is tapered down to a diameter of approximately $1\,\mu \rm{m}$ over a region of 20\,mm, using a hydrogen-flame brushing technique. A second identical aspherical lens is used for coupling back to free-space and the probe field is then interfered with the local oscillator and guided into a balanced homodyne detector. The differential homodyne signal is spectrally analysed using an ESA. The individual dc components from the homodyne detectors are fed to a PI-controller, actively stabilizing the local oscillator phase via a piezo actuated mirror. Optical coupling to the toroidal resonator is controlled by positioning the micro toroid with respect to the tapered fibre, using a piezo actuated 3-axis stage, thereby changing the overlap of the evanescent taper and toroid fields.

\begin{figure}[h]
\centerline{\includegraphics{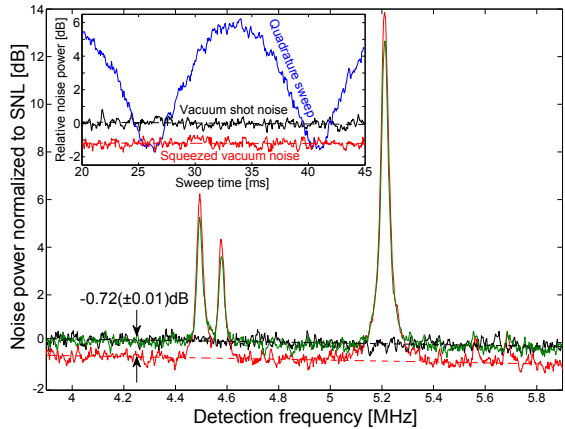}}
\caption{\label{fig_MechSpectrum} Optomechanically transduced vibration spectrum, measured without (green) and with squeezed light (red), yielding a transduction sensitivity reduced below the SNL (black). All traces are averaged over 10 samples, recorded with RBW = 10\,kHz and VBW = 100\,Hz. \newline
Inset: Characterisation of the transmitted vacuum squeezed state. All traces were recorded in zero span mode at a detection frequency of 4.9\,MHz, RBW = 100\,kHz, and VBW = 100\,Hz. The detector dark noise (25\,dB below shot noise) has not been subtracted.}
\end{figure}

To characterize the squeezed state transmitted through the tapered fibre, we decouple the toroid and perform homodyne measurements of the field quadratures $X(\theta) = X_1 \cos(\theta) + X_2 \sin(\theta)$ by linearly sweeping the LO phase. The recorded data are presented in the inset of  Fig.\ref{fig_MechSpectrum}. Linear fits to vacuum and squeezed noise levels yields   $-1.20(\pm 0.03)$\,dB of squeezing, corresponding to a squeezing factor $r=0.138(\pm 0.003)$, limited by large coupling and propagation losses in the tapered fibre amounting to $\gtrsim 30\%$. We attribute the majority of the losses to adhesion of dust~\cite{Fujiwara2011} and water to the fibre, causing increased scattering from the tapered region. Formation of fractures in the glass due to tensile stress~\cite{Brambilla2006}, is also a possible contributor to the optical losses. The homodyne visibility was 98($\pm 1$)\% and the quantum efficiency of the homodyne detector photodiodes (Epitaxx ETX-500) is 87($\pm 2$)\%.

We establish optical coupling to the toroid by bringing it in the proximity of the tapered fibre, staying in the highly under-coupled regime. A combination of coarse thermal tuning of the optical cavity resonances, using a Peltier element underneath the toroid, and a fine frequency tuning of the main laser, is used to bring the probe light on resonance with an optical toroid mode with $\kappa/2\pi = 180\,\rm{MHz}$ (FWHM). Fig.~\ref{fig_MechSpectrum} shows the recorded spectral noise power of the balanced homodyne signal with the LO phase locked for phase quadrature detection, in the case of coherent (green) and phase squeezed (red) probe states.
Transduced signals from three mechanical vibration modes are visible in the spectrum, and most importantly the background noise floor is reduced by $-0.72(\pm 0.01)$\,dB below SNL when squeezed light is injected. The reduction was evaluated by linear fits to the recorded noise levels (dashed lines). All traces were recorded with probe carrier and local oscillator powers of $20\,\mu\rm{W}$ and $1.2\,\rm{mW}$, respectively.  The variation in peak levels for the two probing schemes is attributed to different taper-toroid coupling strengths.

In conclusion, we have experimentally demonstrated a quantum-enhancement of the optomechanical transduction sensitivity using squeezed states of light to probe the mechanical vibrations of a tapered fibre coupled micro toroidal resonator. A transduction sensitivity $-0.72\,\rm{dB}$ below SNL was achieved, primarily limited by losses in the tapered fibre. We are confident that a 3\,dB improvement of the mechanical transduction sensitivity is attainable through further optimization of the tapered fibre production, shielding and nitrogen-purging of the taper-toroid region, and increased homodyne detection efficiency by use of high-quantum efficiency diodes. Reaching this level of impact would render our quantum-enhanced transduction scheme feasible for practical implementations. One immediate application would be to improve the efficiency of optoelectromechanical feedback cooling, where the cooling limit is set by the signal-to-noise ratio of the transduction signal~\cite{Lee2010}.
It is important to note that the demonstrated quantum-enhancement technique not only applies to cavity optomechanics in the under-coupled regime. In general, the best transduction sensitivity is achieved at critical coupling, but since $\Omega_m /\kappa \ll 1$ in the discussed system, working in the under-coupled regime is favorable, in order to minimize degradation of squeezing at relevant frequencies. However, a sideband-resolved system with $\Omega_m/\kappa >1$~\cite{Schliesser2009b} can in principle be operated in the critical-coupling regime with negligible loss of squeezing, yielding an optimal sensitivity enhancement.
\\ \\
This research was funded by the Danish Council for Independent Research (Sapere Aude program) and the Lundbeck Foundation.

\end{document}